\documentclass[amsmath,amssymb,showkeys,aps,pra,reprint,footnoteinbib]{revtex4-2} 
\usepackage{amsmath,amssymb}
\usepackage[english]{babel}
\usepackage{ bbold }
\usepackage{upgreek}
\usepackage{dsfont}
\usepackage{graphicx}
\usepackage{comment}
\usepackage{braket}
\usepackage{colortbl}
\usepackage{blindtext}
\usepackage{hyperref}
\begin{document}

\title{Direct and clean loading of nanoparticles into optical traps at millibar pressures}

\author{Maryam Nikkhou$^{\dagger*}$, Yanhui Hu$^{\dagger}$, James A. Sabin and James Millen}

\thanks{$^{\dagger}$ These authors contributed equally to this work}
\thanks{\\$*$ Correspondence: maryam.nikkhou@kcl.ac.uk; \\ james.millen@kcl.ac.uk.}

\affiliation{Department of Physics, King's College London, Strand, London, WC2R 2LS, United Kingdom}
    
\begin{abstract}
Nanoparticles levitated by optical fields under vacuum conditions have applications in quantum science, the study of nanothermodynamics and precision sensing. Existing techniques for loading optical traps require ambient conditions, and often involve dispersion in liquids, which can contaminate delicate optics and lead to enhanced optical absorption and heating. Here we present a clean, dry and generic mechanism for directly loading optical traps at pressures down to 1\,mbar, exploiting Laser Induced Acoustic Desorption. Our method allows rapid and efficient trapping, and is also suitable for site-selective loading of nanofabricated particles grown on a silicon substrate.
\end{abstract}

\maketitle

\section{Introduction}
A nanoparticle levitated in high vacuum ~\cite{Millen2020} provides an ideal platform for testing macroscopic quantum mechanics ~\cite{MillenQ2020} and for studies of nano-thermodynamics ~\cite{Millen2018}. Levitation offers extreme isolation of the nanoparticle from the thermal environment, making them excellent low dissipation mechanical devices for sensing force ~\cite{Ranjit2016}, pressure ~\cite{Kuhn2017} and torque ~\cite{Ahn2020}. \par

While there are various nanoparticle levitation techniques ~\cite{Millen2020} the use of focussed optical fields offers high precision and control, and has enabled cooling and control at the quantum level ~\cite{Aspelmeyer2014,Delic2020}. Precision applications require trapping under vacuum conditions, however due to the conservative nature of optical potentials a dissipation mechanism is required to load nanoparticles into optical traps. \par

A common loading technique is to suspend nanoparticles in a solvent, and then introduce them into the gas phase through nebulization in ambient conditions. The particles then diffuse to the optical trap and are captured, typically after a few minutes ~\cite{Summers2008,Delic2019}, after which the pressure is reduced. This method has some drawbacks, namely the undesirability of bringing high-vacuum systems repeatedly to ambient pressures, the potential to contaminate delicate optics in the trapping region, and absorption of the solvent into porous dielectric nanoparticles ~\cite{Delic2019}. Nebulization also relies on having a large number of nanoparticles available, since the method is highly probabilistic, and so isn't suitable for small samples of tailored particles. \par

\begin{figure}[!ht]
\centering
\includegraphics[width=8.5cm]{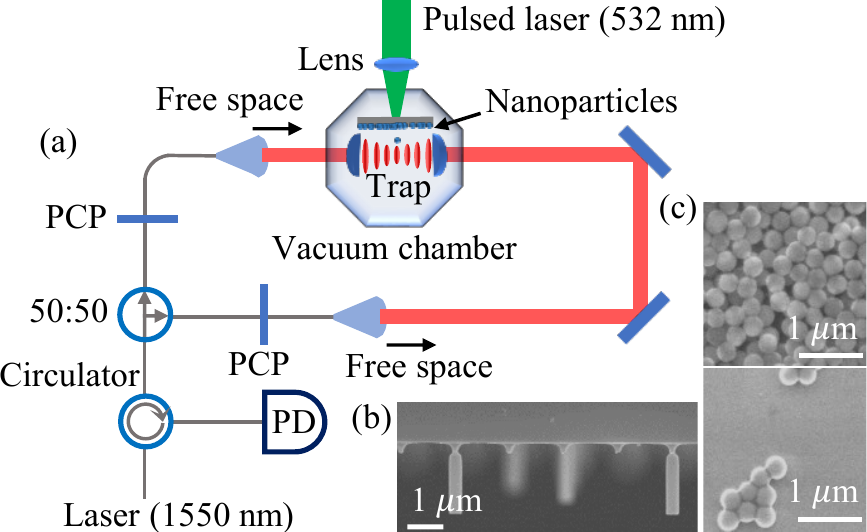}
\caption{(a) \textbf{Experimental setup}. Nanoparticles are trapped in a counter-propagating standing-wave trap formed by two beams of equal power and linear polariztion, at a wavelength of $1550\,$nm. The trap is formed using aspheric lenses with the working distance of $2.1\,$mm yielding a beam waist of $6\,\mu$m. Unless otherwise stated, the total optical power in the trap is 200\,mW. The particles are launched downwards using LIAD from a substrate located 8\,mm above the optical trap. The particle’s motion is encoded in the light which travels back through the optical system and is directed to a photodetector (PD) by an optical circulator. Polarization controlling paddles (PCPs) are used to modify the polarization of the light. (b) SEM micrograph of nanorods grown on a silicon wafer, with missing particles illustrating single-particle launching. (c) SEM micrograph of nanospheres deposited on a substrate before (top) and after (bottom) launch.}
\label{Figure1} 
\end{figure}

Another method is to launch particles from the surface of a piezoelectric transducer ~\cite{Ashkin1971,Arita2013,Millen2015}. This method is clean and dry, and works at pressures down to a few mbar. Launching requires enough mechanical force to overcome the van der Waals force between the piezoelectric surface and the particles, making this method suitable only for particles with diameters greater than $\sim 400\,$nm. As with nebulization, piezoelectric launching isn't suitable for nanofabricated particles grown on a substrate. \par

In this work, we use Laser-Induced Acoustic Desorption (LIAD), a dry and vacuum compatible method for loading nanoparticles into optical traps. In the LIAD method, a pulsed laser beam is focused onto the back side of a substrate upon which particles are distributed or grown. The pulse generates acoustic shock-waves through thermo-mechanical stress to locally eject particles from the substrate. LIAD has been used for launching biological cells ~\cite{Peng2007}, strands of DNA ~\cite{Bald2008} and silicon nanoparticles ~\cite{Asenbaum2013}. This method can overcome the Van der Walls force between the particles coated onto a surface, and can also launch 3D etched silicon nanoparticles directly from their silicon substrate ~\cite{Millen2016}. The particles are only ejected from the region of the laser focus, allowing selective and efficient launching, and LIAD works for dielectric particles from $<100\,$nm up to several micrometres at pressures down to 1\,mbar. \par

Here, we will characterize the LIAD method as a technique for loading nanoparticles into an optical trap, presenting an efficient, flexible, clean and vacuum-compatible tool for the field of levitated optomechanics.

\section{Experimental method}

In this manuscript, we will describe and characterize a method for launching commercial silica nanospheres from a substrate, since these are widely used by researchers. We note that we also use this method to launch nanofabricated silicon nanorods (Supplied by Kelvin Nanotechnology, length: 950\,nm, diameter 200\,nm, grown with a separation of $10\,\mu$m, with an underetched breaking-point, following the work presented in \cite{Kuhn2015}.) ~\cite{Millen2016}, where it is possible to launch individual nanoparticles, as illustrated in Fig. \ref{Figure1}(b). 

Silica nanospheres (Non-functionalized dry silica supplied by Bangs Labs) with diameters of $300\,$nm are dispersed in an isopropanol solution and sonicated in an ultrasonic bath for an hour to prevent aggregation. The dispersion of nanoparticles is pipetted onto a $1\,$cm$^2$ of aluminium sheet of thickness $400\,\mu$m. Rigid aluminium sheet as a substrate has been shown to produce a lower mean launch velocity than aluminium foil, silicon wafer, or titanium foil ~\cite{Millen2016}. After drying, the sample is placed inside the vacuum chamber 8\,mm above the optical trap, with the nanoparticles facing towards the trap, as shown in Fig. \ref{Figure1}(a). \par

Our optical trap is formed within the vacuum system by two counter-propagating focused laser beams (NKT Photonics Koheras AdjustiK seed laser with BoostiK E15 amplifier) with a wavelength of $1550\,$nm, where each beam possesses an identical polarization. The trapping light is back-coupled through the optical system, and separated by a circulator onto a photodiode (Fig.\ref{Figure1}(a)). This signal is maximized to ensure optimal alignment of the trapping beams, and provides read-out of trapped nanoparticle motion. \par

We focus a pulsed laser beam (Litron Lasers NANO-S 120-20) onto the backside of the aluminium substrate to create an acoustic shock, causing release of the nanoparticles via LIAD. The pulsed laser has a wavelength of 532\,nm, a pulse length of $\sim 4.6\,$ns, and we typically operate it in single-shot mode with a peak intensity of 588\,GW/cm$^2$. The waist of the laser focus on the backside of the substrate is $\sim 17\,\mu$m.\par

From the front side of the substrate, the nanoparticles are launched towards the optical trap, well directed along the Poynting vector of the pulsed laser. Careful alignment of the launch laser relative to the optical trap centre is vital. Due to the intensity profile of the focussed launch laser beam, the launch area is typically smaller than the area described by the beam waist.  Fig.\ref{Figure1}(c) shows a scanning electron micrograph (SEM) of a substrate coated by nanospheres before and after launch. 

\subsection{Velocity distribution}

\begin{figure}[!t]
\centering
\includegraphics[width=8.0cm]{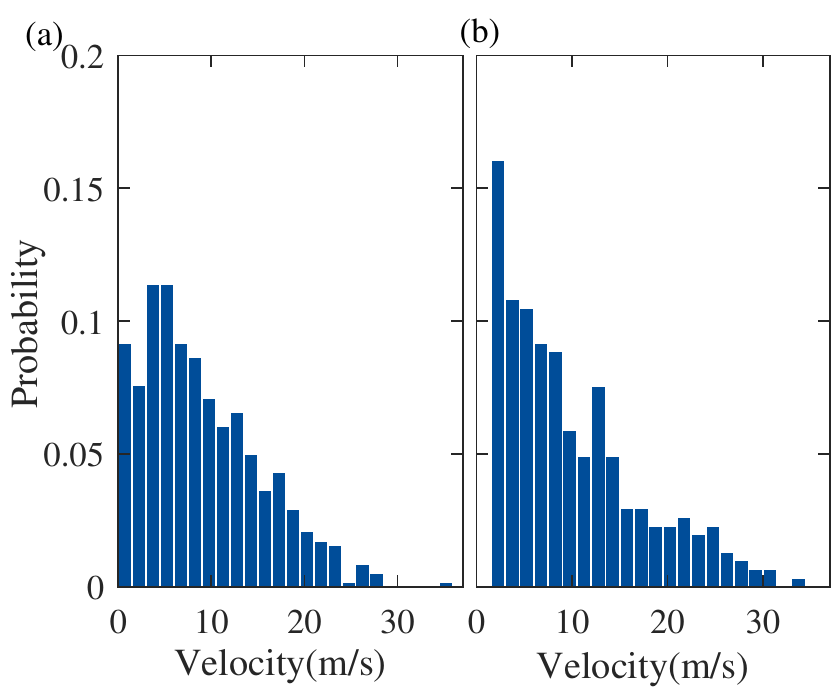}
\caption{\textbf{Velocity distributions} for 300\,nm diameter silica nanospheres launched from a 400\,$\mu$m thick aluminium sheet at a pressure of (a) $2.5\times 10^{-7}$\,mbar and (b) 0.12\,mbar. Histograms are reconstructed from over 120 events for each pressure.}
\label{Figure2}
\end{figure}

We briefly present the velocity distribution of the launched particles, which is analyzed in more detail in ref. ~\cite{Millen2016}. In that previous work it was found that the velocity distribution is independent of the pulsed laser intensity, and we will not consider the physical mechanism of LIAD further in this manuscript. 

To study the velocity distribution of nanospheres in our experiment, the particles are launched using laser pulses with a peak intensity of 588\,GW/cm$^2$, and their arrival time at the optical trap is recorded, as presented in Fig.\ref{Figure2}. Gravity plays an insignificant role over the observed timescales.  \par

Due to the finite transit-time of the particles from launch until observation, the gas pressure plays a significant role in determining the velocity distribution at the optical trap. By working at a low pressure of $2.5\times 10^{-7}$\,mbar (where it is not possible to directly load the trap) the mean-free-path of the nanoparticle between collisions with gas molecules is much longer than the transit path. Hence, the velocity distribution at the optical trap, as presented in Fig.\ref{Figure2}(a), can be considered an accurate representation of the initial launch velocity distribution.  \par

At a higher pressure of 0.12\,mbar, the nanoparticle will go through thousands of collisions before reaching the trap region. This lowers the mean velocity, as presented in Fig.\ref{Figure2}(b), but also removes the slowest particles since they diffuse and never reach the optical trap. 

\begin{figure*}[ht]
	 {\includegraphics[width=0.65\textwidth]{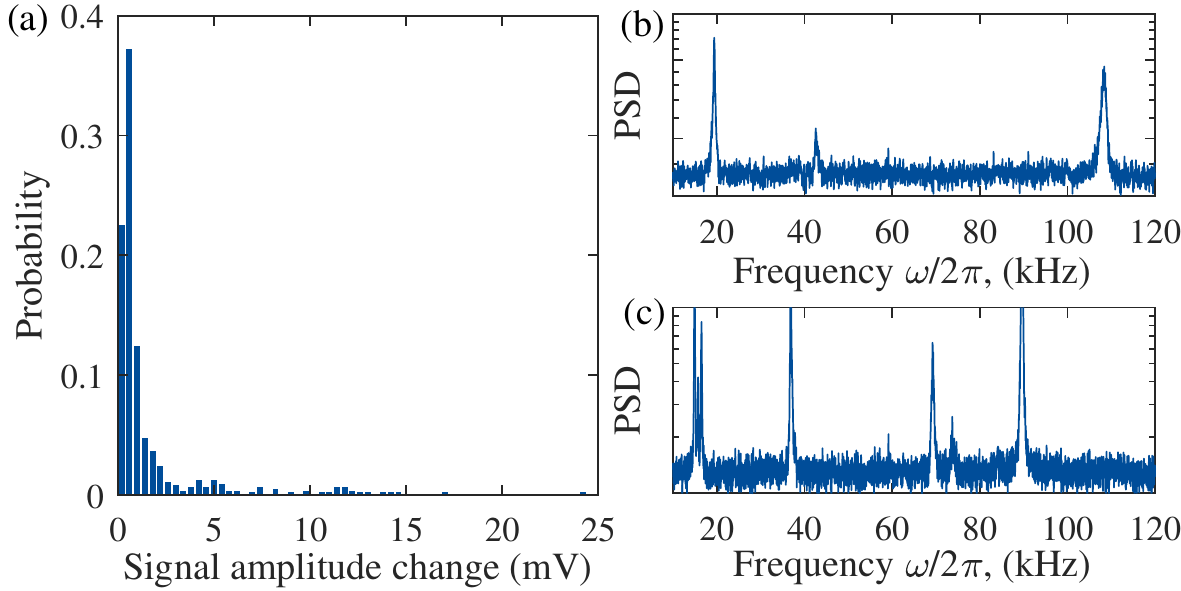}}	
\caption{\textbf{Identifying and characterizing trapping events} (a) Histogram of the signal change due to the trapping events. We identify a peak corresponding to single trapped particles, and a long tail due to clusters and multiple trapped particles. (b) Power Spectral Density (PSD) of the center-of-mass motion of a single particle or cluster. (c) PSD corresponding to multiple trapped particles.}
\label{Figure3}
\end{figure*}

\subsection{Identifying trapping events}

In most applications it is required to trap a single nanoparticle, and in this manuscript we describe the optimal conditions for achieving this. There are three catagories of possible trapping events; single particle trapping, single particle-cluster trapping, and multiple separated particle or cluster trapping. Trapping events cause a change in our detected signal amplitude due to light scattering, and particle motion can be subsequently tracked. \par

Figure \ref{Figure3}(a) shows a histogram of the signal amplitude change in response to a trapping event, reconstructed from over 800 events. We observe a peak corresponding to trapping single particles, and a long tail corresponding to clusters or multiple particles. The width of the single-particle peak is due to the variability in exact trapping location within the optical standing wave. \par

The signal can be analyzed to reconstruct the power spectral density (PSD) of trapped particle motion, as shown in Fig. \ref{Figure3}(b), yielding frequencies corresponding to centre-of-mass motion. Trapping multiple particles or clusters yields a more complex PSD, as shown in Fig. \ref{Figure3}(c). Hence, by analyzing the overall amplitude change of our signal and its PSD, we can identify trapping events corresponding to single trapped particles. \par

\section{Results}

We use LIAD to load nanoparticles directly into an optical trap at various pressures. Interactions between the nanoparticles and residual gas are required to provide the dissipative mechanism to load the conservative optical trap. However, unlike previously mentioned nebulization techniques, our loading method does not over-rely on the diffusion of nanoparticles into the trapping region, as they are directed by the LIAD mechanism, dramatically increasing the efficiency of trap loading.  

\subsection{Trapping efficiency with pressure}

\begin{figure}[!t]
\centering
\includegraphics[width=8.5cm]{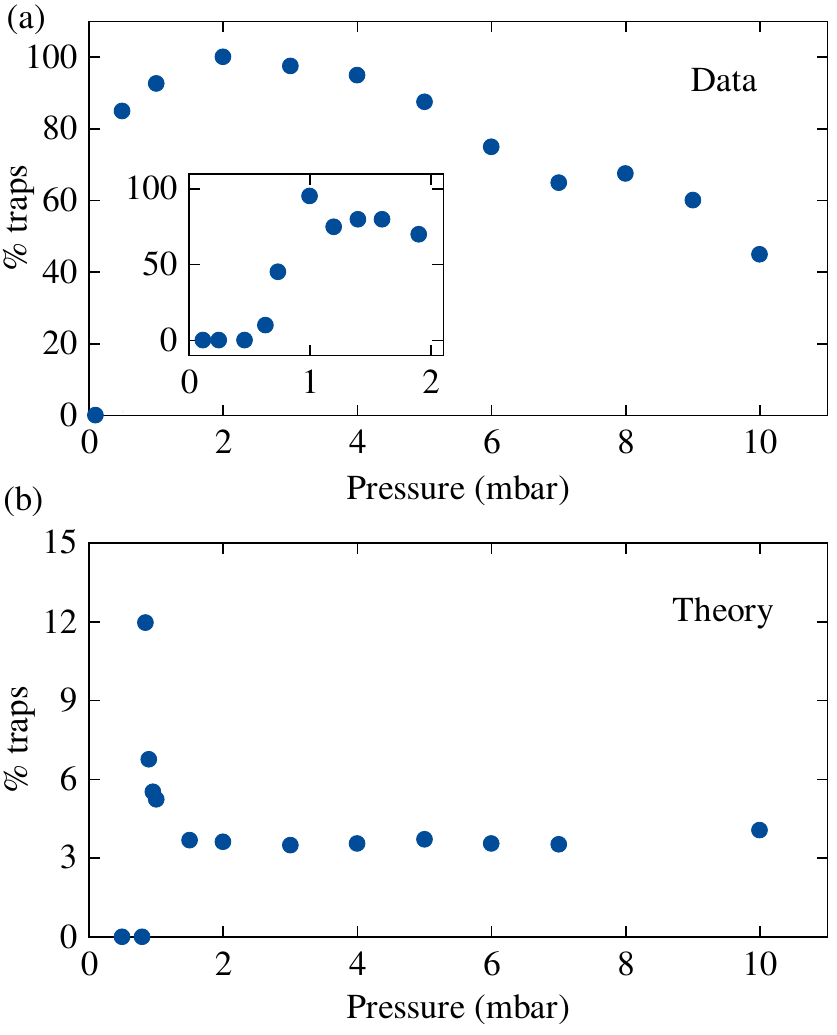}
\caption{\textbf{Dependence of loading efficiency upon pressure.} (a) The probability of trapping events as a function of residual gas pressure, with a loading laser intensity of 588\,GW/cm$^2$. The inset shows results of a separate experimental run focussed around the optimal trapping pressure of 1\,mbar. Each datapoint represents 20 LIAD pulses. (b) Theoretical simulation of optical trap loading, predicting the optimal loading efficiency at about 1\,mbar. Each point represents 10,000 launching events. Note that experimentally multiple particles are launched per shot, explaining the higher efficiency than theoretically predicted.}
\label{Figure4}
\end{figure}

We find that our trap loading technique is highly efficient across a wide range of presssures, as shown in Fig. \ref{Figure4}(a). At 1\,mbar we are able to load single particles into the optical trap with $>80\%$ probability per LIAD pulse.   \par

Once particles are released from the substrate via LIAD, they travel towards the trap region in a direction defined by the Poynting vector of the launch laser, continuously losing energy through collisions with gas molecules. Eventually they slow to their terminal velocity, at which point they fall under gravity and diffuse. The equation describing the vertical position of the particles, neglecting diffusion, is:

\begin{equation}
y(t) = \left(1-e^{-\Gamma t}\right)\left(\frac{g}{\Gamma^2} + \frac{u}{\Gamma}\right) - \frac{g}{\Gamma}t,
\end{equation}
where $\Gamma$ is the momentum damping rate (see ~\cite{Millen2018} for definitions), $g$ is the acceleration due to gravity and $u$ is the initial velocity, which in our case is negative. The mean-square-displacement of the particle in any one dimension due to diffusion is $\langle (\Delta s)^2\rangle = 2k_BT/(M\Gamma) t$, where $T$ is the ambient gas temperature, and $M$ is the mass of the nanosphere ~\cite{Millen2018}. For guidance on stochastic simulation of particle motion, see for example ~\cite{Volpe2013}.

By constructing a stochastic simulation of a particle launched towards an optical potential, we model the efficiency of our trap loading process, as shown in Fig. \ref{Figure4}(b). If the pressure is too low, then the particles pass through the trap region without slowing, and are not trapped, leading to a sharp pressure threshold in terms of the minimum trapping pressure. There is a critical pressure at which the particles reach their terminal velocity in the trapping region, yielding enhanced capture probability. Above this pressure, particles reach the trapping region by diffusion, and the trapping probability levels off. Under these conditions the particles may take many hundreds of seconds to be trapped.  \par

We compare our simulation to the experimental data shown in Fig. \ref{Figure4}(a). The inset shows that we observe a sharp turn on in trapping probability, and an optimal trapping pressure of about 1~mbar, as predicted. The data is less sharp than the simulation, since in the simulation the particles are all given the same initial velocity, whereas in the experiment the particles have the velocity distribution as presented in Fig. \ref{Figure2}. Experimentally, the trapping efficiency decreases with increasing pressure, which we believe is due to experimental runs being terminated too early, before the hundreds of seconds it can take for trapping to occur.

\subsection{Launch laser intensity}

\begin{figure}[!t]
\centering
\includegraphics[width=8.5cm]{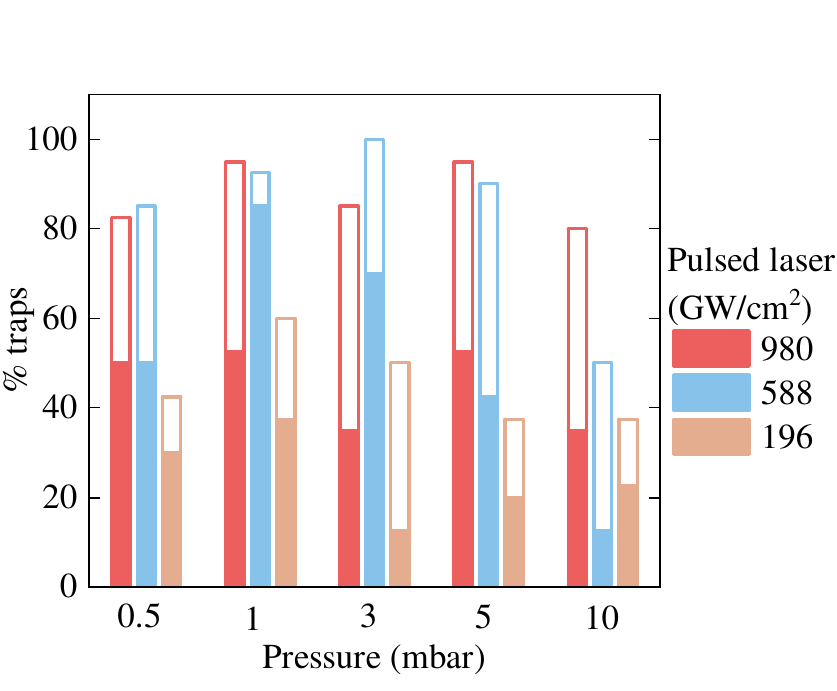}
\caption{\textbf{Effect of launch laser intensity and pressure on the efficiency of trapping single particles.} Solid bars represent single trapped particles, and empty bars represent clusters or multiple particles, as identified using the methods described in Fig. \ref{Figure3}. Low launch laser intensities yield few particles, whereas high intensities lead to multiple trapping events per shot. Each bar represents 40 LIAD pulses.}
\label{Figure5}
\end{figure}

The effect of the launch laser intensity on the LIAD process is complex, and surprisingly doesn't significantly alter the launch velocity above a minimum threshold, as discussed elsewhere ~\cite{Millen2016}. However, the launch laser intensity strongly effects the number of particles launched per shot, and the effective area from which the particles are launched. \par

These points are illustrated in  Fig. \ref{Figure5}. Each colour represents a different launch laser intensity. When the intensity is low, the trapping efficiency drops as fewer particles are removed from the substrate. The solid bars in the histogram represent single particle trapping events, and the empty bars represent trapping of clusters or multiple particles. At higher launch laser intensities we see that the overall trapping probability is not significantly increased, but the chance of trapping multiple particles increases.  \par

\subsection{Optical trap depth}

\begin{figure}[!t]
\centering
\includegraphics[width=8.5cm]{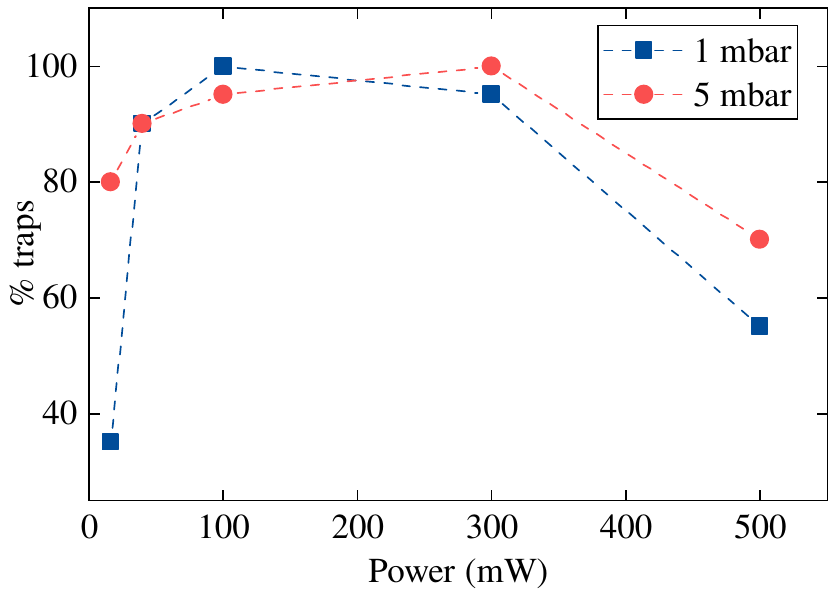}
\caption{\textbf{Effect of optical trap power on trapping efficiency}, for two different pressures. The optical trap power must be high enough to be able to confine a particle that is in thermal equilibrium with the environment. Each data point represents 40 LIAD pulses.}
\label{Figure6}
\end{figure}

The optical trap is formed by two counter-propagating tightly focused laser beams, as described above. We choose a counter-propagating trap over single-beam trap to create a larger trapping volume to maximize the likelihood of the particle trapping. We show the effect of optical trap power in Fig. \ref{Figure6}. The optical trap must be at least deep enough to confine a particle in thermal equilibrium with the environment, and realistically it must be significantly deeper due to thermal fluctuations. This yields a lower threshold for the trap power required to capture a particle. In our experiment we are able to trap with optical powers as low as $\sim 10$\,mW. It is easier to trap at low powers at higher pressures, due to the additional dissipation provided by the gas. \par

We also observe that the trapping efficiency drops at high optical trap power, which we attribute to particle absorption and heating, which is further supported by increased stability at higher pressures, where residual gas can cool the particles through collision ~\cite{Millen2014}.

\section{Conclusions}
We have presented a clean, dry and efficient method for loading single nanoparticles into optical traps, using Laser Induced Acoustic Desorption (LIAD). We find that there is an optimum pressure at which the traps can be loaded, which will depend on the size of particle and the distance between the sample and the trap. We find that by tuning the launch laser intensity, we can enhance the probability of trapping single particles. Around the optimum pressure the particle motion is near ballistic and trapping occurs in a few milliseconds, and at higher pressures particles diffuse into the trapping region. We are also able to launch particles from localized regions of a sample, enabling the launch and trapping of sparse particles. \par

We note that Bykov et al. \cite{Bykov2019} developed a technique combining LIAD and the temporal control of a Paul trap potential to launch and capture charged nanoparticles directly under ultrahigh vacuum (UHV) conditions. This technique does not require a dissipation mechanism, since the potential is turned-on when the particle is at the centre of the trap, at which point it does not have enough energy to escape. However, since in general \textit{optical} potentials are much less deep than those in a Paul trap, direct UHV loading into an optical trap would only work for nanoparticles with velocities $<<0.1\,$ms$^{-1}$, requiring the development of new soft-launching techniques.

\section*{Acknowledgments}
The authors would like to thank Ben Blackburn at the King's College London Physics Research Facility for assistance with SEM imaging. J.M. acknowledges P. Asenbaum, S. Kuhn and M. Arndt at the University of Vienna for the initial conception of the LIAD technique for launching nanoparticles.

\end{document}